\documentclass[crcready,onecolumn]{iosart2x}
\makeatletter
\let\numberlines@hook\relax
\makeatother

\usepackage{dcolumn}

\newcolumntype{d}[1]{D{.}{.}{#1}}

\firstpage{1}
\lastpage{5}
\volume{1}
\pubyear{0000}

\begin{document}
\begin{frontmatter} 

%
\title{Simulating neutron transport in long beamlines at a spallation neutron
source using Geant4}

\runningtitle{Proceedings of ICANS-XXIII}

\author[A,B]{\fnms{Douglas D.} \snm{DiJulio}\ead[label=e1]{douglas.dijulio@esss.se}%
\thanks{Corresponding author. \printead{e1}.}}
\author[A,B]{\fnms{Isak} \snm{Svensson}}
\author[A,C]{\fnms{Xiao Xiao} \snm{Cai}}
\author[B]{\fnms{Joakim} \snm{Cederkall}}
and
\author[A]{\fnms{Phillip M.} \snm{Bentley}}
\address[A]{European Spallation Source ESS ERIC, SE-221 00 Lund, Sweden}
\address[B]{Department of Physics, Lund University, SE-221 00 Lund, Sweden}
\address[C]{DTU Nutech, Technical University of Denmark, Roskilde, Denmark}

\begin{abstract}
The transport of neutrons in long beamlines at spallation neutron sources
presents a unique challenge for Monte-Carlo transport calculations. This is due to the
need to accurately model the deep-penetration of high-energy neutrons through meters of
thick dense shields close to the source and at the same time to model the transport of low-
energy neutrons across distances up to around 150 m in length. Typically, such types of calculations
may be carried out with MCNP-based codes or
alternatively PHITS. However, in recent years there has been an increased interest in
the suitability of Geant4 for such types of calculations. Therefore, we have implemented
supermirror physics, a neutron chopper module and the duct-source variance reduction technique for low-
energy neutron transport from the PHITS Monte-Carlo code into Geant4. In the current work, we present a series of
benchmarks of these extensions with the
PHITS software, which demonstrates the suitability of Geant4 for simulating long neutron beamlines
at a spallation neutron source, such as the European Spallation Source, currently under construction
in Lund, Sweden.
\end{abstract}

\begin{keyword}
\kwd{Geant4}
\kwd{PHITS}
\kwd{Spallation Source}
\kwd{Neutron supermirror}
\end{keyword}
\end{frontmatter}

\section{Introduction}
At modern high-power spallation neutron sources, a wide-range of neutron energies (below meVs up to GeVs) are generated
through the bombardment of a heavy metal target with a high-energy proton beam (up to a few GeV). High-energy neutrons
which escape the target volume are deeply penetrating and thus spallation targets are typically surrounded by several meters
of thick shields containing metal and concrete. Low-energy neutrons, which are used for scientific investigations, are created
through the placement of moderators next to the target position. After undergoing collisions in the moderator material, the
low energy neutrons (meV and below) enter beamports, placed around the moderator position, and are then transported to the
sample position of a neutron scattering instrument via neutron guides. These instruments can be up to about 150 m in length at modern
spallation sources, such as the European Spallation Source (ESS \cite{Garoby18}), currently under construction in Lund, Sweden. The combination of thick
dense shields near the target position, where attenuation is driven largely by nuclear processes, with the need to
transport low-energy neutrons over great distances, via reflection/diffraction processes, presents a unique challenge for
Monte-Carlo neutron transport calculations. \\
\indent Exasperating this challenge is that typical Monte-Carlo software used to simulate the proton beam interactions in the spallation target
do not come packaged with the ability to simulate the transport of low-energy neutrons via coherent scattering.
Currently, the Particle and Heavy Ion Transport code System (PHITS \cite{PHITS}) does come packaged with this
feature while other popular codes such as the Monte-Carlo N-Particle Transport (MCNP) family of codes \cite{MCNP},
the Geant4 Toolkit \cite{Geant4,Geant4Ref1}, and FLUKA \cite{FLUKA} do not. This deficiency has been addressed in a number of
different ways, which include for example patched versions of
MCNPX \cite{MCNPX,Gallmeier2009} and MCNP6 \cite{MCNP6,Magan} or interfacing
with neutron-ray tracing software such as McStas \cite{McStas,McStas2} or
Vitess \cite{Vitess1,Vitess2} using specialized tools \cite{Klinkby2013,Knudsen2014,Klinkby2014,Kittelmann2017,Brydevall2015}.
In addition, the PHITS software also includes a powerful technique for enhancing the efficiency of the supermirror transport,
referred to as the duct-source method. The details of this method will be further discussed below. \\
\indent The majority of these developments have focused on PHITS and the MCNP-based codes while Geant4 has seen
little application in neutron optics and shielding calculations at spallation neutron sources. There
has however been a recent push towards enhancing the low-energy neutron transport features of Geant4,
driven largely by developments in neutron detector design \cite{Kittelmann2014,Kittelmann2015}.
This has resulted in an increased interest in understanding the suitability of Geant4 for neutron optics
and shielding calculations at ESS \cite{Cherkashyna13,Cherkashyna14,Cherkashyna15,Stenander2015,DiJulio2016ess,DiJulio2016}.
Another reason why it is interesting to explore the suitability of Geant4 for these types of calculations
is that the software is freely available online and open source. Thus new developments can be immediately
put to use by other scientists around the world and$/$or further expanded upon if desired. \\
\indent In the current work, we have implemented supermirror capabilities, a neutron chopper module and the duct-source variance
reduction from PHITS into the Geant4 toolkit. Furthermore, we also discuss how to extend the duct-source technique for neutron deep-shielding
problems. In the following, we first present an overview of the these developments and their implementation in Geant4 and then show some results of calculations using these methods.
The work will focus on the Geant4 developments, as the PHITS work has been described in detail in \cite{Takeda}.
\section{Overview of the techniques}
\begin{figure}[t]
\includegraphics[width=110mm]{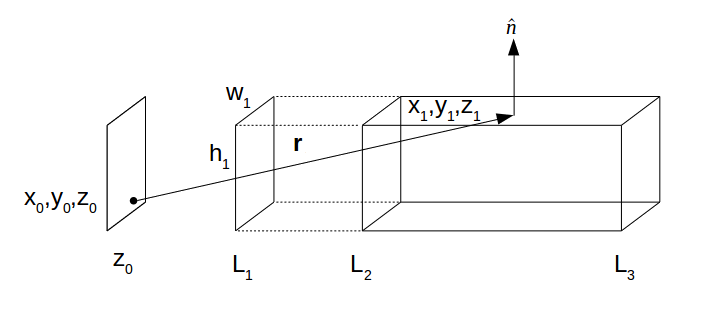}
\caption{An illustration of the duct-source technique. See the text for the detailed description of the figure.}
\end{figure}
\subsection{Duct-source variance reduction}
Due to the unique long-pulse nature of the ESS, a significant number of the neutron beamlines will be around 150 m in length. The current of neutrons crossing a shielding wall
at 150 m will be around five orders of magnitude smaller than at 2 m from the moderator position. To diminish this variance during a Monte-Carlo simulation,
the duct-source method was developed and implemented into PHITS. The basic idea of the method
is to generate neutrons uniformly along the walls of an imaginary duct-source region that overlaps with the physical beamline geometry.
The weights of the neutrons are adjusted accordingly to the distance from the source region. 
In this way, the statistics are completely uniform
across the entire length of the duct-source, improving dramatically the efficiency of a given Monte-Carlo calculation at distances far from the neutron source location. \\
\indent An illustration of the duct-source technique is shown in Fig. 1. Neutrons which originate at the position $x_0$, $y_0$, $z_0$ are generated
uniformly on the surfaces of the virtual duct, of width $w_1$, height $h_1$, starting at $L_1$ and ending at $L_3$. The physical geometry
of the beamline starts at $L_2$. The weight of a neutron generated on this surface is given by,
\begin{equation}
  w = \frac{A}{4 \pi r^{2}} \cdot \frac{\mathbf{r}}{\left\lVert{\mathbf{r}}\right\rVert} \cdot \hat{\mathbf{n}}, \\
\end{equation}
where $A$ is the total area (side walls and the end wall combined) of the duct-source imaginary walls, $r$ is the distance from the source,
and $\hat{\mathbf{n}}$ is the normal vector to the respective wall \cite{Svensson}. \\
\indent An important parameter in the implementation is the fraction of particles, $PT$, that pass through the end of the duct. This parameter
can be controlled by the user and results in the following two adjusted weights calculated from Eq. (1),
\begin{equation}
  w_s' = \frac{P_s}{1-PT}w, \\
\end{equation}
for particles passing through the sides of the duct and
\begin{equation}
  w_e' = \frac{P_e}{PT}w, \\
\end{equation}
\begin{figure}[t]
\includegraphics[width=110mm]{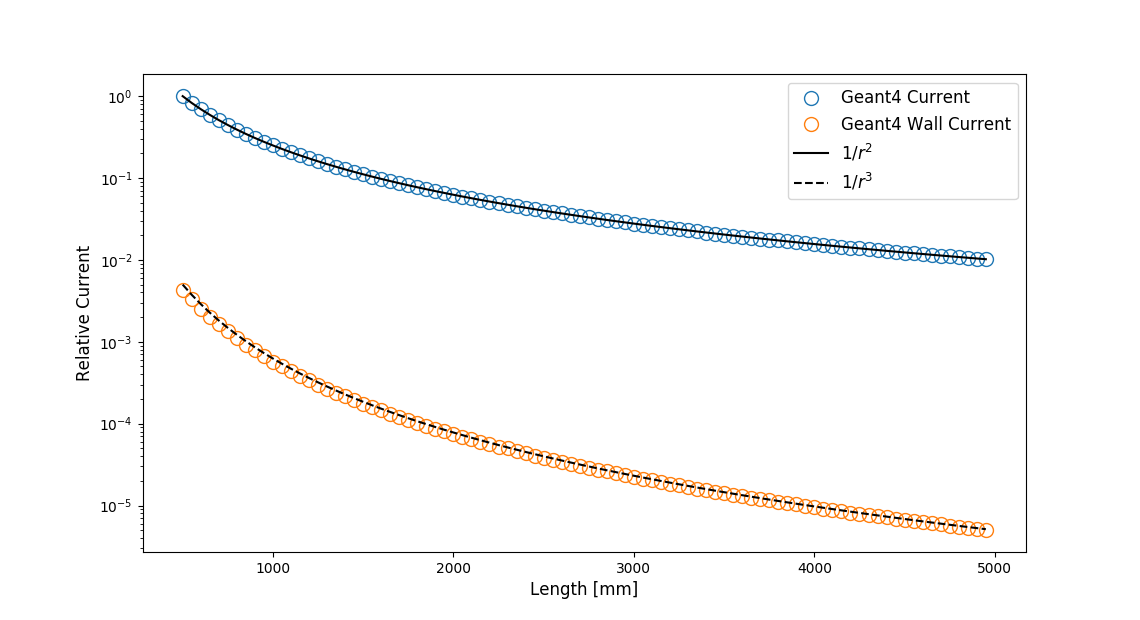}
\caption{The current inside a 50 m guide and the wall current from Geant4, using the duct-source technique, plotted with the trendlines $1/r^2$ and $1/r^3$.}
\end{figure}
for particles passing through the end of the duct, where $P_s$ represents the fraction of the total area representing the sides of the duct surface
and $P_e$ is the fraction of the total area representing the end of the duct surface. In the case that a neutron passes through the end of the duct,
the value $r$ is calculated from the intersection of a neutrons trajectory with the wall of an infinitely long duct-source region. \\
\indent Our implementation of the duct-source also allows for a non-uniform angular distribution in the angle, $\theta$, relative to the
neutron beam direction pointing down the guide. The angle, $\phi$, which describes the direction perpendicular to the beam direction, is considered
to be uniform. This description is typical for source terms used at spallation neutron sources \cite{Gallmeier2006}. The procedure is carried out
by adjusting the weights in Eqs. (2-3) by
\begin{equation}
  w_a = \frac{P_i}{\Delta\mu}w_{2,3}, \\
\end{equation} 
where $P_i$ represents the fraction of the total neutron flux in the angular bin $\Delta\mu$, $w_{2,3}$ is the weight calculated from either Eq. (2) or (3),
and $\mu$ is the cosine of the angle $\theta$. This implementation makes it possible to capture both
the highly an-isotropic fast neutron current entering the guide and the isotropic like low-energy current entering the guide, which are typical
of spallation neutron sources. Additionally, we have also introduced the possibility for energy source biasing in a similar manner as Eq. (4) in our implementation, in order
to emphasize the relatively low, but highly-important, fraction of high-energy neutrons in a typical spallation source spectrum. \\
\indent The above description has been coded in a Geant4 program as a PrimaryGeneratorAction class.
\subsection{Supermirror physics}
The supermirror reflectivity function as described in PHITS is given by,
\begin{equation}
\begin{array}{l}
R = R_0 \qquad Q \le Q_c,  \\
R = \frac{1}{2}R_0(1-$tanh$[(Q-mQ_c)/W])(1-\alpha(Q-Q_c)) \qquad Q > Q_c,         
\end{array}
\end{equation}
where $Q_c$ is the material critical scattering vector, $Q$ is the scattering vector, $\alpha$ describes the reflectivity decline, $m$ is the m-value parameter and $W$ is the cutoff width.
This function has been implemented into Geant4 as a biasing process. The biasing process is defined as a surface between two volumes. Whenever a neutron of the appropriate energy
crosses the surface, it is split into two particles, one with a weight multiplied by $R$ and one with a weight multiplied by 1-$R$. This makes it possible to both simulate the neutron
transport down the guide and also the losses into the shielding outside of the guide. The detailed interactions of the low-energy neutrons inside of the supermirror coating
is not included \cite{Rodion}.
\subsection{Chopper module}
We have implemented the neutron chopper module using a biasing process in a similar manner to the supermirror physics implementation. The geometry, material and dynamic properties
of the chopper are defined in the Geant4 program source code. During the simulation, when a neutron is incident on the front surface of the chopper, the time of its arrival
is compared to the dynamic properties of the chopper system. If the chopper should be open, the neutron is transported to the back surface of the chopper, as if it did not
have an interaction within the chopper material. If the chopper should be closed, the neutron is allowed to be transported through the chopper material. 
\section{Results}
The results presented in this section were generated using Geant4 version 10.3 and the QGSP$\_$BERT$\_$HP physics list. The PHITS calculations were carried out using
PHITS 2.88 and the prepackaged nuclear libraries that come with PHITS, as described in \cite{Takeda}. \\
\begin{figure}[t]
\includegraphics[width=160mm]{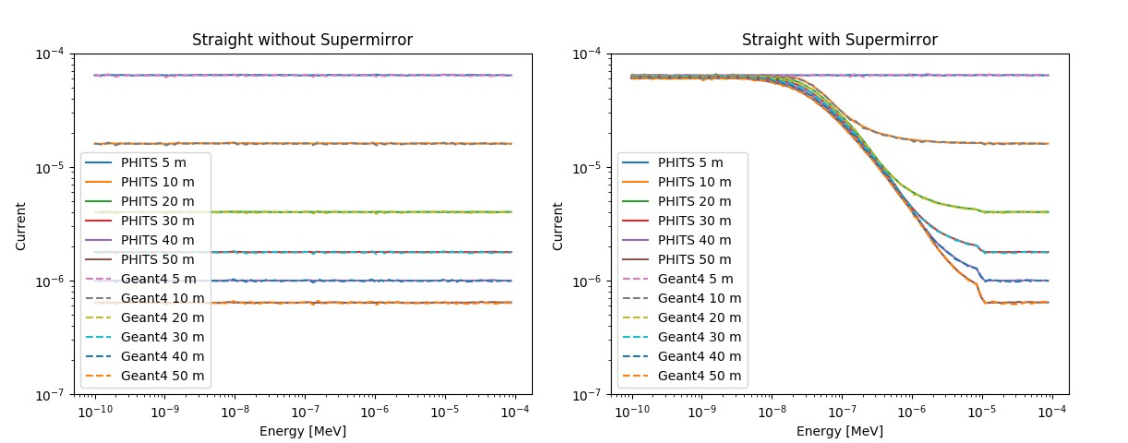}
\caption{(Left) Current inside a 50 m long guide (left) without the supermirror physics activated and (right) with the supermirror physics activated. The PHITS calculations are fron \cite{Takeda}. }
\end{figure}
\indent The duct-source calculated current inside a 50 m long guide and the current on the walls is shown in Fig. 2. It can be noted that the current on the walls of the
guide falls off as $1/r^{3}$ and the current inside the guide falls off as $1/r^{2}$. The trendlines are shown in the plot and validate the usage of the duct-source technique.  \\
\indent A comparison between PHITS and Geant4 calculations using the supermirror option is shown in Fig. 3 for a 50 m long straight guide. The m-value of the guide was set to $m=3$. The energy spectrum used as input
to the calculations was flat across the entire energy range. In the left panel, the supermirror physics was turned off, and results indicate the $1/r^{2}$ fall off in the guide.
In the right panel, the supermirror physics were turned on and the transport of the low-energy neutrons from the source to 50 m can be clearly seen. At higher energies, the fall off
as $1/r^{2}$ can be seen again. The agreement between PHITS and Geant4 is excellent. \\
\indent The neutron current along the entire length of a straight 50 m guide is plotted in 2D in Fig. 4. In the top panel, the supermirror physics were turned off and in the bottom
they were turned on. One can clearly see the divergence of the transported neutrons at the end of the guide, as shown in the bottom panel. \\
\begin{figure}[t]
\includegraphics[width=140mm]{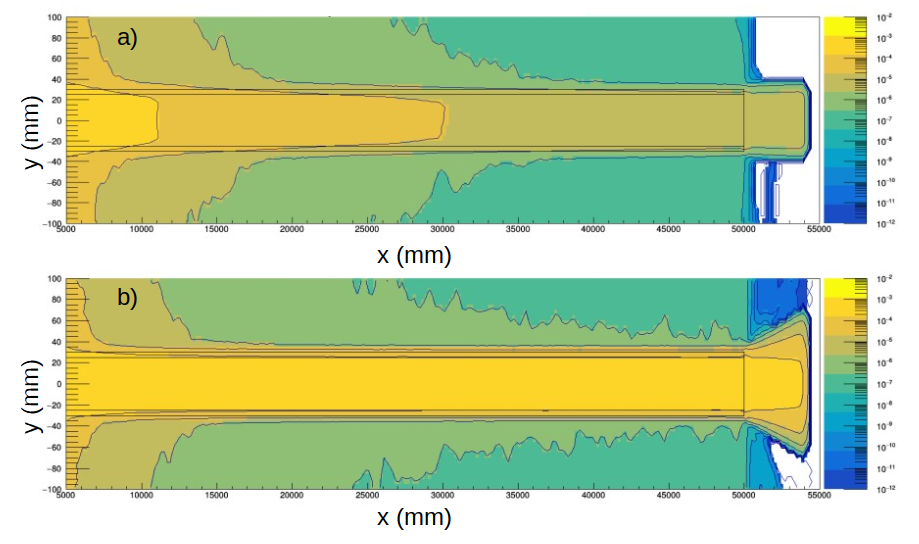}
\caption{Relative neutron current along the entire length of straight a guide with a) the supermirror physics turned off and with b) the supermirror physics turned on.}
\end{figure}
\begin{figure}[t]
\includegraphics[width=160mm]{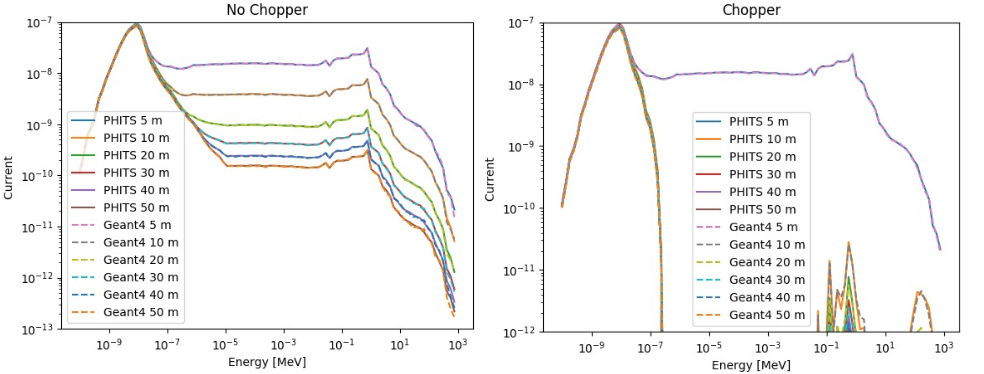}
\caption{Comparison between PHITS and Geant4 calculations for a (Left) 50 m straight guide with a neutron chopper completely open and (Right) with a neutron chopper turned on. The PHITS calculations are fron \cite{Takeda}. }
\end{figure}
\indent The comparisons for a 50 m guide with a neutron chopper not activated and activated are shown in Fig 5. The length of the chopper blade was 30 cm and consists of Inconel, as described in \cite{Takeda}.
The supermirror transport and duct-source option were turned on in both cases, as can be seen
in the transport of the low-energy neutrons across the full 50 m. When the chopper was activated, the low-energy neutron peak remains, as the chopper was open at these times, and
the high-energy current was significantly reduced. It can be noted that a small fraction of the high-energy neutrons still made it through the chopper blade material, which
can further be avoided if the guide system was curved out of line-of-sight of the source. It can be again seen that the agreement between Geant4 and PHITS is excellent.
\section{Conclusion}
In summary, we have developed three extensions for Geant4, including the duct-source technique, supermirror physics and a neutron chopper module. We have carried out
several benchmark tests between the Geant4 implementations and the PHITS code system, revealing excellent agreement between the two. We are currently continuing to debug
some minor differences in our implementations and have also preliminary demonstrated that the duct-source technique can be readily extended to sample uniformly across curved surfaces,
at least up to the point that the guide is outside line-of-sight of the source. Future further tests could be the simulation of more realistic beam geometries and
also the combination with other variance reduction techniques, as carried out in Ref. \cite{Svensson}.

\section{Acknowledgments}
D.D.D. would like to thank T. Miller for presenting this work at the ICANS conference in his absence. D.D.D. would
also like to thank Masatoshi Arai for his inovlement in the PHITS/Geant4 benchmarking study.

\end{document}